\documentclass{amsart}
\usepackage{latexsym}
\usepackage{amsmath}
\usepackage{amssymb}
\usepackage{amscd}
\usepackage{bm}
\newtheorem{Th}{Theorem}[section]
\newtheorem{Cor}{Corollary}[section]
\newtheorem{Prop}{Proposition}[section]
\newtheorem{Lem}{Lemma}[section]
\newcounter{Remark}[section]
\newcounter{Example}[section]

\newcommand{\bet}{\begin{Th}}
\newcommand{\ent}{\stepcounter{Cor}
   \stepcounter{Prop}\stepcounter{Lem}
   \stepcounter{Remark}\stepcounter{Example}\end{Th}}

\newcommand{\bec}{\begin{Cor}}
\newcommand{\enc}{\stepcounter{Th}
   \stepcounter{Prop}\stepcounter{Lem}
   \stepcounter{Remark}\stepcounter{Example}\end{Cor}}
\newcommand{\bep}{\begin{Prop}}
\newcommand{\enp}{\stepcounter{Th}
   \stepcounter{Cor}\stepcounter{Lem}
   \stepcounter{Remark}\stepcounter{Example}\end{Prop}}
\newcommand{\bel}{\begin{Lem}}
\newcommand{\enl}{\stepcounter{Th}
   \stepcounter{Cor}\stepcounter{Prop}
   \stepcounter{Remark}\stepcounter{Example}\end{Lem}}

\newcommand{\Remark}{
   \stepcounter{Remark}
   \noindent{\bf Remark\,  \thesection.\theRemark \, }:
   \stepcounter{Th}\stepcounter{Cor}\stepcounter{Prop}
   \stepcounter{Lem}\stepcounter{Example}}

\newcommand{\Proof}{\noindent{\it Proof\,}:\ }

\newcommand{\R}{{\mathbb{R}}}
\newcommand{\C}{{\mathbb {C}}}
\newcommand{\Z}{{\mathbb {Z}}}

\newcommand{\ga}{\gamma}
\newcommand{\pa}{\partial}

\newcommand{\La}{\Lambda}
\newcommand{\la}{\lambda}
\newcommand{\Si}{\Sigma}
\newcommand{\si}{\sigma}

\begin{document}
\def\RR{\mathbb{R}}
\def\CC{\mathbb{C}}
\def\HH{\mathbb{H}}
\def\vp{\varphi}
\def\Sig{\Sigma}
\def\om{\omega}
\def\al{\alpha}
\def\be{\beta}
\def\si{\sigma}
\def\Si{\Sigma}
\def\pa{\partial}
\def\tr{\mbox{\rm tr}}
\def\bpa{\bar\partial}
\def\bi{\bar i}
\def\ga{\gamma}
\def\Ga{\Gamma}
\def\eps{\epsilon}
\def\s{\sqrt}
\def\Ca{\mathcal{C}}
\def\ad{\mbox{\rm ad}}
\def\Ad{\mbox{\rm Ad}}
\def\triad{\triangledown}
\def\bpmat{\begin{pmatrix}}
\def\epmat{\end{pmatrix}}
\def\beq{\begin{equation}}
\def\eeq{\end{equation}}
\def\P{\mathbb{P}}
\def\Q{\mathbb{Q}}
\def\F{\mathbb{F}}

\title [The Bryant-Salamon $G_2$-manifolds and hypersurface geometry]
{The Bryant-Salamon $G_2$-manifolds and hypersurface geometry}
\author{R. Miyaoka}
\address{Graduate School of Mathematics, Kyushu University, Fukuoka, 812-8581/JAPAN}
\email{r-miyaok@math.kyushu-u.ac.jp}
\thanks{The author is partially supported by
Grant-in-Aid, 16204007, Ministry of Education and Science of Japan}


\begin{abstract}
We show that two of the Bryant-Salamon $G_2$-manifolds have 
a simple topology, $S^7\setminus S^3$ or $S^7\setminus\C P^2$.
In this connection, we show there exists a complete Ricci-flat 
(non-flat) metric
on $S^n\setminus S^m$ for some $n-1>m$. We also give many
 examples of special Lagrangian submanifolds 
of $T^*S^n$ with the Stenzel metric. 
Hypersurface geometry is essential for these arguments.
\end{abstract}

\maketitle

\section{Introduction}
\label{intro}

Physicists as well as mathematicians are interested in Ricci-flat 
(K\"ahler) manifolds as a special case of Einstein manifolds.
Ricci-flat metrics are often constructed on vector bundles
over Riemannian manifolds where
some group action of cohomogeneity one is effectively used.
In this case, the base manifold is regarded as a
degenerate orbit, while the principal orbits are of 
codimension one in the total space, i.e., the sphere bundles.

This reminds us of the theory of isoparametric hypersurfaces. 
In fact, isoparametric hypersurfaces in the sphere $S^n$
exist in one parameter families, which laminate $S^n$  
with two other degenerate submanifolds, called the focal submanifolds.
If we delete from $S^n$ one of the focal submanifolds, then
 the remaining part is  a disk bundle
over another focal submanifold \cite{Mu}.
Two of the Bryant-Salamon $G_2$-manifolds fit this theory exactly.  
Namely, the spin bundle $\mathcal{S}$ over $S^3$ is associated to the isoparametric hypersurfaces $S^3\times S^3$ in $S^7$, 
and the anti-self-dual bundle 
$\La_-^2(\C P^2)$ over $\C P^2$ is 
to the so-called Cartan hypersurfaces in $S^7$.
Using these, we prove:

\bet
There exist homeomorphisms $\mathcal{S}\cong S^7\setminus S^3$ 
and $\La_-^2(\C P^2)\cong S^7\setminus \C P^2$, 
where $S^3$ and $\C P^2$ are embedded in $S^7$ in the 
standard way. In other words, a compactification 
of $\mathcal{S}$ and $\La_-^2(\C P^2)$ is given by $S^7$.
\label{top}
\ent

On the topology of $\mathcal{S}$ and $\La_-^2(M)$, some details 
are described in \cite{AW}, but our viewpoint from 
the theory of isoparametric hypersurfaces is new. 
Moreover, this theorem reminds us of the Calabi-Yau problem on open
manifolds, 
\cite{BK}, \cite{TY}, 
which asks if there exists a complete Ricci-flat K\"ahler 
metric on the complement of some divisor $D$ of a compact K\"ahler
 manifold $\overline{M}$ 
with positive Ricci curvature. It seems natural to pose a real
version of this problem:

\medskip\noindent
{\bf Problem.} When can we construct a complete Ricci-flat, 
non-flat metric on the complement of some subset $D$ of a compact 
irreducible Riamannian manifold $\overline{M}$ with positive Ricci curvature?

\medskip
In the case $\overline{M}=S^n$, we give 
 a partial answer (\S\ref{homo}).  
\bep A complete non-flat Ricci flat metric exists on 
\begin{enumerate}
\item $S^7\setminus \C P^2$, $S^7\setminus S^3$ : the Bryant-Salamon metric
\item $S^{m+2n+2}\setminus S^{2n+1}$, $n,m\geq 1$ : the L\"u-Page-Pop metric 
\item $S^6\setminus S^2$, $S^{14}\setminus S^6$ : the Stenzel metric 
\end{enumerate}
\label{rf}
\enp

Another aspect of Ricci flat manifolds is a relation with 
special geometry. With respect to the Stenzel metric, 
 $T^*S^n$ becomes a Calabi-Yau manifold with calibrations
$\Re(e^{i\theta}\Omega)$, where $\Omega$ is the global 
holomorphic $n$-form (see \S\ref{Stenzel}).
Then what are special Lagrangian submanifolds?
Here again the theory of isoparametric hypersurfaces works well.
Harvey and Lawson's result \cite{HL}, 
generalized by Karigiannis and Min-Oo
recently \cite {KM}, tells us that the conormal bundle over an austere 
submanifold in $S^n$ is a special Lagrangian submanifold of 
$T^*S^n$.
We have many examples of austere submanifolds of $S^n$ in \cite{IKM}, 
hence we obtain

\bet The conormal bundles of the focal
submanifolds $W_\pm$ of {\em any} isoparametric hypersurfaces 
are special Lagrangian submanifold of $T^*S^n$  equipped with 
the Stenzel metric. 
Infinitely many non-homogeneous examples are included among them. 
Moreover, 
the conormal bundle of the following {\em minimal} isoparametric hypersurfaces in 
$S^n$ are special Lagrangian submanifold of $T^*S^n$, 
$$
\begin{array}{ll}
W=S^{n-1}\\
W^{2d}= S^d\times S^d, \quad n=2d+1\\
W^{3d},\quad n=3d+1\quad d=1,2,4,8\\
W^{4d},\quad n=4d+1\quad d=1,2\\
W^{6d},\quad n=6d+1,\quad d=1,2
\end{array}
$$
where $W$ has, respectively, $1,2,3,4,6$ principal curvatures. 
The phase $e^{i\theta}$ is determined by the dimension of 
$W_\pm$ or $W$.
\label{SL}
\ent

\Remark
In \cite{IKM}, we prove that the conormal bundles of the cones 
in $\R^{n+1}$ over all above $W_\pm$ or $W$
are special Lagrangian cones in $\C^{n+1}$.

\section{The Bryant-Salamon $G_2$-manifolds}
\label{BSmetric}

By a $G_2$-manifold, we mean a Riemannian manifold with 
the holonomy group $G_2$.  The metric is called a $G_2$-metric.

Denoting the exterior product of three vectors of an orthonormal coframe
$e^1,\dots,e^7$ of $\R^7$  by $e^{ijk}=e^i\wedge e^j\wedge e^k$, 
define a 3-form $\phi$ on $\R^7$ by
\begin{equation}
\phi=e^{125}-e^{345}+e^{136}-e^{426}+e^{147}-e^{237}+e^{567}.
\label{3f}
\end{equation}
The automorphism group $G_2$ of the Cayley numbers can be defined also as
the subgroup of $GL(7,\R)$ preserving $\phi$ \cite{B1}. 
A $G_2$-structure on a 7-dimensional manifold $X$ is 
a reduction of the structure group of the linear frame bundle to $G_2$.
Let $\mathcal{O}$ be the $GL(7,\R)$-orbit of $\phi$ ($\cong GL(7,\R)/G_2$), 
then a $G_2$-structure is equivalent to the existence of a global 3-form
$\phi$ on $X$ satisfying $\phi_x\in \mathcal{O}_x$. 
Since $G_2\subset SO(7)$, a $G_2$-structure induces a Riemannian 
metric $g$ on $X$. The holonomy group 
is contained in $G_2$ if and only if 
$d\phi=d*\phi=0$, and is equal to $G_2$ if and only if 
 there are no non-trivial parallel 1-forms on $X$,
provided that $X$ is simply connected and connected. 
Note that a  $G_2$-metric is Ricci flat \cite{B1}.

The first examples of $G_2$-manifolds were given by 
Bryant \cite{B1}.
Later on, complete $G_2$ metrics were constructed 
by Bryant and Salamon \cite{BS} on the spin bundle $\mathcal{S}$ 
over $S^3$ and on the anti-self dual 
bundle $\La_-^2(M)$ of the self-dual Einstein manifolds 
$M=S^4$ and $\C P^2$.
Compact Riemannian manifolds with holonomy $G_2$ were constructed 
by Joyce \cite{J} via a generalization of the Kummer construction,
and by Kovalev \cite{K} via a twisted gluing method.
In both cases, Ricci flat metrics on non-compact 
manifolds are essential tools. 
Indeed, in the former case, $G_2$-metrics which are 
asymptotically locally Euclidean were used to 
connect two non-compact parts, and in the latter,
$SU(3)$-metrics obtained by solving the open Calabi-Yau problem \cite{TY} 
played an important role.
However,  the latter two constructions do not give metrics explicitly.
 
The Bryant-Salamon metrics are explicit. Indeed, 
let $X=\mathcal{S}, \La_-^2(S^4)$ or $\La_-^2(\C P^2)$. 
Let $g_b$ and $g_f$ be the standard metrics on the base and 
the fiber space, respectively,  normalized appropriately by constant
multiples. 
Consider a hypersurface $N_r$ of $X$ consisting of 
fiber vectors of length $r$.
Now, on $X=\cup_{r\ge 0}N_r$, 
they seek a metric 
so that the 3-form which depends on the metric 
satisfies the non-linear partial differential equations 
$d\phi=d*\phi=0$.
Restricting the metric
to the warped product form $g=u(r)f_b+v(r)g_f$, 
where $u(r)$ and $v(r)$ are functions of $r$,
they reduce the equations to ODE's, and 
obtain $G_2$-metrics 
$$
\begin{array}{ll}
g=(\lambda+r^2)^{2/3}g_b+(\lambda+r^2)^{-1/3}g_f\quad \mbox{\rm on } \mathcal{S}\\
g=(\lambda+r^2)^{1/2}g_b+(\lambda+r^2)^{-1/2}g_f\quad \mbox{\rm on } \La_-^2(M^4).
\end{array}
$$

Actually, when $\lambda>0$, these metrics extend to 
complete ones on 
$X=\cup_{r\ge 0}N_r$. The non-existence of non-trivial parallel 
1-forms is then proved, which establishes Hol$(g)=G_2$.
It turns out that a homogeneous metric
is used on $N_r$, where $N_r\cong S^3\times S^3$ for 
$X=\mathcal{S}$, $N_r\cong \C P^3$ for $X=\La_-^2(S^4)$,
and $N_r\cong SU(3)/T^2$ ($T^2$ is a maximal torus) 
for $X=\La_-^2(\C P^2)$.
We note that, however, the metric 
is different from the standard one. Indeed 
when $X=\La_-^2(S^4)$, the metric on $N_r\cong \C P^3$
is not the Fubini-Study metric, but a non-K\"ahler Einstein metric.
The one for $N_r$ in the case  $X=\La_-^2(\C P^2)$ is also  non-K\"ahler 
Einstein.

We notice here that $S^3\times S^3$ and the flag manifold 
$SU(3)/T^2$ are homeomorphic to  
typical isoparametric hypersurfaces in $S^7$.
In particular, the latter is called the Cartan hypersurface,
on which the induced metric is K\"ahler Einstein.


\section{Homogeneous and isoparametric hypersurfaces in the sphere}
\label{homo}

Now, we give a brief review of 
homogeneous and isoparametric  hypersurfaces in the sphere.

By isoparametric hypersurfaces, 
we mean hypersurfaces with constant principal 
curvatures (see \cite{Th}). 
These are given as level sets $W_t=f^{-1}(t)$ of the so-called 
Cartan-M\"unzner function $f:S^n\to [-1,1]$, for $t\in (-1,1)$. 
The level sets $W_{\pm}=f^{-1}(\pm 1)$ have lower dimension and 
are called the focal submanifolds. 
The function $f$ is extended to $F:\R^{n+1}\to \R$ so that $f=F|_{S^n}$, 
where $F$ is a homogeneous polynomial of 
degree equal to the number of the principal curvatures, 
satisfing two PDE's, see \cite{Mu}. 
Note that $W_{\pm}$ and $W_0$ are minimal submanifolds. 
The most important fact in our argument is that  the ambient sphere is 
stratified as 
\begin{equation}
S^n=\cup_{t\in [-1,1]}W_t,
\label{str}
\end{equation}
by hypersurfaces $W_t$, $t\in (-1,1)$ and the
two focal submanifolds $W_{\pm}$. 

Typical examples of isoparametric hypersurfaces are given by
homogeneous hypersurfaces. 
Let $G/K$ be an $(n+1)$-dimensional rank two symmetric space of 
compact type, and let 
${\frak g}={\frak k}+{\frak m}$ be the associated decomposition,  
where $\frak g$ and $\frak k$ are Lie algebras of $G$ and $K$,
respectively. 
At $o\in G/K$, the tangent space 
$T_o(G/K)$ is identified with $\frak  m\cong \R^{n+1}$, 
equipped with the invariant metric induced from the one on $\frak g$.
Then $K$ acts on $\frak m$ as an isometry by the adjoint action, 
which we call the isotropy action. 
Let $S^n$ be the unit sphere of $\frak m$.
Then the principal orbit of $K$ through $x\in S^n$ is a hypersurface 
$N=(\mbox{\rm Ad} K)x$ of $S^n$, because the rank of $G/K$  is two. 
Note that we obtain a one parameter 
family of such hypersurfaces, and the two singular orbits called the 
focal submanifolds. 
Conversely, 
every homogeneous hypersurface in a sphere is obtained in this way \cite{HsL}, and all such hypersurfaces are classified. 

\medskip
On the other hand, there exist infinitely many non-homogeneous 
isoparametric hypersurfaces with four principal curvatures. 
These were constructed by Ozeki and Takeuchi \cite{OT} by using the 
representation of Clifford algebras.
Later on, Ferus, Karcher and M\"unzner generalized the method and 
obtained systematically the so-called isoparametric hypersurfaces
of FKM type \cite{FKM}.
The only known examples of non-homogeneous isoparametric
hypersurfaces are of this type.

\medskip
\Remark Recently, isoparametric hypersurfaces with four principal curvatures are classified by Cecil, Chi and Jensen \cite{CCJ} 
except for 10 cases, and 
they are either of FKM type or homogeneous.

\medskip
The number $g$ of principal curvatures is limited to 1,2,3,4, or 6, and 
of particular interest is in the case of 3 and 6, where orbits 
of a subgroup of $G_2$ appear \cite{Mi}. 
Recall that those with three principal curvatures are known as 
Cartan hypersurfaces, which are tubes over standard embedded 
Veronese surfaces $\F P^2$ where $\F=\R,\C,\HH,  
{\mathcal C}$ in $S^4,S^7,S^{13}$ and $S^{25}$  
(${\mathcal C}$ is the Cayley algebra) \cite{C}. 
 Veronese surfaces are related to the so-called  Severi varieties \cite{AB}. 

We are concerned with the case $S^7$.  
All isoparametric hypersurfaces in $S^7$ are homogeneous and so are classified.
We denote a $k$-dimensional sphere with radius $a$ by $S^k(a)$.
Let $\bm x=(x_1,\dots,x_8)$ be the coordinate of $\R^8$.
\begin{enumerate}
\item[(a)] $g=1$ : $F(\bm x)=x_8$, $f^{-1}(t)=S^6(\sqrt{1-t^2})$, $t\in (-1,1)$. $W_\pm=$ north and south poles. 
\item[(b)]
$g=2$ : $F(\bm x)=\sum_{i=1}^{k+1}x_i^2-\sum_{j=k+2}^8x_j^2$, 
$f^{-1}(t)=S^k(\sqrt{\frac{1-t}{2}})\times S^{6-k}(\sqrt{\frac{1+t}{2}})$, $t\in (-1,1)$,  $1\leq k\leq 5$ :
 generalized Clifford torus. $W_\pm=S^k(1)$ and $S^{6-k}(1)$.
\item[(c)] 
$g=3$ : $F(\bm x)=u^3-3uv^2+\dfrac32u(|x|^2+|y|^2-2|z|^2)+\dfrac{3\sqrt3}{2}v(|x|^2-|y|^2)+\dfrac{3\sqrt3}{2}(xyz+\bar x\bar y\bar z)$,  
$\bm x=(u,v,x,y,z)\in \R^2\times \C^3=\R^8$. 
 $f^{-1}(t)$ $\cong SU(3)/T^2$ : Cartan hypersurface$=$ 
isotropy orbits of 
$SU(3)\times S(3)/SU(3)$.  $W_\pm=$ two copies of $\C P^2$.
\item[(d)]
$g=4$ :  $F(\bm x)$ is a polynomial of degree 4 (see \cite{OT}). $f^{-1}(t)=$ isotropy orbits of $SO(6)/SO(2)\times SO(4)$.
\item[(e)]
$g=6$ :  $F(\bm x)$ is a polynomial of degree 6 (see \cite{OT}). 
$f^{-1}(t)\cong SO(4)/\Z_2=$ isotropy orbits of $G_2/SO(4)$.
\end{enumerate}

\medskip\noindent
{\em Proof of Theorem \ref{top} and of Proposition \ref{rf} (1)}

Consider a hypersurface $N_r$ of 
$\mathcal{S}$ consisting of 
the fiber vectors of length $0<r<\infty$.  
Then $N_r$ is homeomorphic to ${S}^3\times S^3(r)$, 
where ${S}^3$ denotes the base manifold. 
Let $W_t$  be as in (b) where $k=3$. 
Now we define
$$
\phi:\mathcal{S} \to S^7\setminus S^3
$$
by identifying ${S}^3$ with $W_-\cong S^3$, 
and then $N_r$, $r\in (0,\infty)$  with 
$W_t=S^3(\sqrt{\frac{1-t}{2}})\times S^{3}(\sqrt{\frac{1+t}{2}})$, 
$t\in(-1,1)$, by 
\begin{equation}
t=\frac{r-1}{r+1},
\label{tr}
\end{equation}
where we keep the correspondence of ${S}^3$ with the first 
$S^3(\sqrt{\frac{1-t}{2}})$ by homothety. 
The continuity at $r=0$ follows because $X$ is a tube over $S^3$. 
This evidently implies that 
$\mathcal{S}\cong\cup_{r\ge 0}{S}^3\times S^3(r)$ is homeomorphic to 
$\cup_{t\in [-1,1)}W_t=S^7\setminus S^3$, 
since $r=\infty$ corresponds to $W_+\cong S^3$. 

The hypersurface $N_r$ of
$\Lambda_-^2(\C P^2)$  is homeomorphic to $SU(3)/T^2\cong W_t$, 
$t\in (-1,1)$, where $W_t=f^{-1}(t)$ in (c). 
Now we define similarly a map, 
$$
\phi:\La_-^2(\C P^2)\to S^7\setminus \C P^2
$$
by identifying the base manifold $\C P^2$ with $W_-\cong\C P^2$, and 
$N_r$, $r\in (0,\infty)$  with $W_t$, $t\in(-1,1)$, by 
(\ref{tr}). The continuity when $r\to 0$ is guaranteed 
since $W_t$ is a tube over $\C P^2$. 
Since $r=\infty$ corresponds to $\C P^2\cong W_+$, 
 we obtain Theorem \ref{top} and Proposition \ref{rf} (1).
\qed

\medskip
\Remark For any isoparametric family, M\"unzner shows that the 
ambient sphere $S^n$ 
can be decomposed into two disk bundles $D_\pm$ over
the focal submanifolds $W_\pm$ so that $S^n=D_+\cup D_-$, 
and $D_+\cap D_-$ is an isoparametric hypersurface, say, $W_0$ 
[M\"u].
If we know this and the stratification (\ref{str}), 
the description $X_1\cup_Y X_2$ (in fact $=S^7$) given in (5.63) of 
\cite{AW} becomes clearer.
\label{remMu}

\medskip
\Remark In the case $M=S^4$, the hypersurface $N_r$ 
of $\La_-^2(S^4)$ is diffeomorphic to $\C P^3$. 
By the result of Cleyton and Swann, \cite{CS}, Theorem 9.3, 
 $\La_-^2(S^4)\cong \C P^3\times \R^1$, 
hence a compactification is given by $\C P^3\times S^1$.

\section{Complete Ricci flat metric on $S^n\setminus D$}

L\"u, Page and Pope constructed complete Ricci flat metrics on 
$S^m\times \R^{2n+2}$ ($m,n\ge 1$), modifying the construction of 
non-homogeneous Einstein metrics on compact manifolds \cite{LPP}.
Since 
$S^m\times \R^{2n+2}\cong \cup_{r\ge 0}S^m\times S^{2n+1}(r)$, 
using the isoparametric embedding of 
$S^m(\sqrt{\frac{1-t}{2}})\times S^{2n+1}(\sqrt{\frac{1+t}{2}})$
into $S^{m+2n+2}$, we see that $S^{m+2n+2}\setminus S^{2n+1}$ 
is a disk bundle over $S^m$, i.e., $S^m\times \R^{2n+2}$. Thus
we obtain (2) of Proposition \ref{rf}. 
(part (3) will be proved in the next section):

\medskip
Note that $S^N\setminus S^n\cong \R^N\setminus \R^n$.
As a less  topologically trivial case, we may ask the following
question which we will discuss on another occasion.

\bep For each isoparametric family $\{W_t\}$ in $S^n$,
does there exist a complete Ricci-flat, non-flat metric 
on $S^n\setminus W_\pm$?
\label{prob11}
\enp

\section{Stenzel metric and calibrated geometry}
\label{Stenzel}

We give a brief introduction to the Stenzel metric on $T^*S^n$. 
Identify $T^*S^n$ with
$\mathcal{Q}^n=\{z\in \C^{n+1}\mid z_0^2+\dots +z_n^2=1\}$
by $T^*S^n\ni (x,\xi)\to x\cosh|\xi|+i\xi/|\xi|\sinh |\xi|$,
and induce a complex structure on $T^*S^n$ from $\mathcal{Q}^n$. 
Then consider a holomorphic $(n,0)$ form $\Omega$ given by
$$
\Omega(T):=(dz_0\wedge \dots \wedge dz_n)(Z,T), \quad
Z=z_0\dfrac{\partial}{\partial z_0}+\dots +
z_n\dfrac{\partial}{\partial z_n}\in \C^{n+1}
$$
The K\"ahler form of the Stenzel metric is given by
$$
\omega_{St}=\dfrac{i}{2}\sum_{j,k=1}^na_{jk}dz_j\wedge d\bar z_k
$$
where
$$
a_{jk}=(\delta_{jk}+\dfrac{z_j\bar z_k}{|z_0|^2})u'+
2\Re(\bar z_jz_k-\dfrac{\bar z_0}{z_0}z_jz_k)u'',
$$
and $u$ is a function of $r=|z|$, of which details we need not here.
This is a highly generalized  Eguchi-Hanson metric, 
first constructed explicitly in \cite{S}.
Note that the Stenzel metric restricted to $S^n$ is the standard
metric on $S^n$. 

\bep There exists a non-flat complete Ricci-flat K\"ahler metric on
$S^6\setminus S^2$ and $S^{14}\setminus S^6$.
\label{s6}
\enp

\Proof Because $S^3$ and  $S^7$ are parallelizable, it follows that 
$T^*S^n\cong S^n\times \R^n\cong\cup_{r\geq 0} S^n\times S^{n-1}(r)
\cong S^{2n}\setminus S^{n-1}$. 
\qed

\medskip
\Remark Stenzel also constructs Ricci-flat K\"ahler metrics on 
the cotangent bundles of rank one symmetric spaces \cite{S}.

\medskip
In the calibrated geometry developed by Harvey and Lawson \cite{HL},
one way to obtain special Lagrangian submanifolds in 
$\C^{n+1}$ is to take the conormal bundle of the so-called
{\em austere} submanifold in $\R^{n+1}$.
A submanifold $N$ in $\R^{n+1}$ or in $S^n$ is austere 
if any shape operator has eigenvalues in pairs $\{\pm \la_j\}$,
and if the multiplicities of $\pm \la_j$ coincide,
where $\la_j=0$ is admissible. 
The cone over an austere submanifold of $S^n$ is austere in $\R^{n+1}$. 
Austere surfaces are nothing but minimal surfaces. 
In some cases, austere submanifolds are 
classified \cite{B2}, \cite{DF}.

In \cite{IKM}, we found  a large class of compact austere 
submanifolds in $S^n$:

\bet \cite{IKM} The focal submanifolds of 
any isoparametric hypersurfaces in $S^n$ are austere. Minimal 
isoparametric hypersurfaces in $S^n$, whose principal curvatures have
the same multiplicity, are austere. 
\label{IKM}
\ent

In fact,  non-zero principal curvatures of minimal isoparametric hypersurfaces appear in $\pm$ pairs, 
and when $m_j$ is the multiplicity of $\lambda_j$,
where $\la_1<\dots<\la_g$, it is known \cite{Mu} that 
\begin{enumerate}
\item[(a)] If $g=2$, then $1\leq m_1\leq m_2=n-m_1-1<n-1$. 
\item[(b)]
If $g=3$, then $m=m_j\in\{1,2,4,8\}$ does not depends on $j$ \cite{C}.
\item[(c)]
If $g=4$, then $m_3=m_1$, $m_4=m_2$, where the pair $(m_1,m_2)$ 
is restricted to those of homogeneous ones or of FKM type \cite{St}.
\item[(d)]
If $g=6$, then $m=1$ or $2$ and $m_j$ does not depend on $j$ \cite{A}.
\end{enumerate}

\noindent
The shape operators of the focal submanifolds have the following eigenvalues:
\begin{enumerate}
\item[(a)] If $g=2$, then $0$. 
\item[(b)]
If $g=3$, then $\pm1/\sqrt3$.
\item[(c)] If $g=4$, then $0, \pm 1$.
\item[(d)]
If $g=6$, then $0$, $\pm1/\sqrt3$, $\pm\sqrt3$.
\end{enumerate}
where the $\pm$ pair of eigenvalues have the same multiplicity.

On the other hand, 
Karigiannis and Min-Oo proved :

\bet \cite{KM}
The conormal bundle of a submanifold $N$ in $S^n$ is a special 
Lagrangian submanifold of $T^*S^n$ with the Stenzel metric  
if and only if $N$ is {austere}.
\label{KM}
\ent

\medskip
From Theorem \ref{IKM} and \ref{KM}, we obtain Theorem \ref{SL}. 

\medskip
\Remark The conormal bundle is given by
$$
\Psi(x(s),\sum t_k\nu^k)=x(s)\cosh |t|+i\hat \nu(s,t)\sinh|t|,
$$
where  $\hat\nu=\sum t_k\nu^k/|t|$,  for $|t|^2=t_{p+1}^2+\dots+t_n^2$,
and an orthonormal frame  $\nu^{p+1},\cdots, \nu^n$ of the conormal space.

\medskip
\Remark When $n=3$, any orientable compact 
(topological) surface
can be minimally immersed in $S^3$ \cite{L},
hence could be austere. Thus the conormal bundles of such surfaces
are special Lagrangian submanifolds of $T^*S^3$.

\medskip
\Remark B.~Palmer shows that the Gauss map
of an isoparametric hypersurface $M$ in  $S^n$ given by  $x\wedge n$,
where $x\in M$ and $n$ is the unit normal,  
defines a {\em Lagrangian} submanifold of Gr$^+_2(n+1,\R)$
 \cite{P}.
Here the oriented 2-plane Grassmanian is identified with 
the complex quadric $Q^{n-1}=\{[z]\in \C P^n\mid z_0^2+\dots +z_n^2=0\}$
equipped with the metric induced from the Fubini-Study 
metric on $\C P^n$. The induced metric has positive Ricci curvature. 
Palmer shows that the Lagrangian submanifold obtained in this way 
is Hamiltonian stable if and only if $M$ is a hypersphere.


\begin{thebibliography}{00}




\bibitem
[1]{A}  U.~Abresch, 
{\em Isoparametric hypersurfaces with four and six principal curvatures}, 
Math. Ann. {\bf  264} (1983), 283--302.

\bibitem[2]{AB}M.~Atiyah and J.~Berndt, 
{\em Projective planes, Severi varieties and spheres},
Surveys in Differential Geometry, Vol. VIII (Boston, MA, 2002), 1--27


\bibitem[4]{AW}
M.~Atiyah and E.~Witten,
{\em M-theory dynamics on a manifold of $G_2$ holonomy},
Adv. Theor. Math. Phys. {\bf 6} (2001), 1--106.

\bibitem[5]{B1} R.~Bryant,
{\em Metrics with exceptional holonomy},
Ann. Math. {\bf 126} (1987), 525--576.

\bibitem[6]{B2} R.~Bryant,
{\em Some remarks on the geometry of austere manifolds},
Bol. Soc. Brasil. Mat. {\bf 21} (1991), 122--157.


\bibitem [7]{BK} S.~Bando and R. Kobayashi,
{\em Ricci-flat K\"ahler metrics on affine algebraic manifolds II}, 
Math. Ann. {\bf 287} (1990), 175--180.

\bibitem[8]{BS}
R. Bryant and S. Salamon,
{\em On the construction of some complete metrics with exceptional holonomy}, 
Duke Math. J. {\bf 58} (1989), 829--250.

\bibitem [9]{C} E.~Cartan, 
{\em Sur des familles remarkables d'hypersurfaces isoparam\'etriques dan
les espaces sph\'erique}, Math. Z. {\bf 45} (1939), 335--367.

\bibitem[10]{CCJ}
T.~Cecil, Q.~S.~Chi and R.~Jensen,
{\em Isoparametric hypersurfaces with four principal curvatures},
to appear in Ann. of Math. (2006).

\bibitem[11]{CS}
R.~Cleyton and A.~Swann, 
{\em Cohomogeneity-one $G_2$-structures},
J. Geom. Phys. {\bf 44} (2002), 202--220.

\bibitem[12]{DF} M.~Dajczer and L.~Florit,
{\em A class of austere submanifolds},
Illinois J. Math. {\bf 45} (2001), 735--755.

\bibitem[13]{FKM}
D.~Ferus, H.~Karcher and H.~M\"unzner,
{\em Cliffordalgebren und neue isoparametrische Hyperfl\"achen},
Math. Z. {\bf 177} (1981), 479--502.


\bibitem[14]{HL} R.~Harvey and H.~B.~Lawson,
{\em Calibrated geometries},
Acta Math. {\bf 148} (1982), 47--157.

 \bibitem [15]{HsL} W.~Y.~Hsiang and H.~B.~Lawson, 
{\em Minimal submanifolds of low cohomogeneity}, 
  J. Diff. Geom. {\bf 5},  (1971), 1--38.

\bibitem [16]{IKM}
G. Ishikawa, M. Kimura and R. Miyaoka,
{\em Submanifolds with degenerate Gauss mappings in the spheres},
 Advanced Stud. in Pure Math. {\bf 37} (2002), 115--149.

\bibitem [17]{J} D.~Joyce,
{\em Compact Manifolds with Special Holonomy}, 
Oxford Sci. Pub. (2000)

\bibitem [18]{K} A.~Kovalev,
{\em Twisted connected sums and special Riemannian holonomy},
J. Reine Angew. Math. {\bf 565} (2003), 125--160.

\bibitem [19]{KM}
S. Karigiannis and M. Min-Oo,
{Calibrated subbundles in non-compact manifolds of special holonomy},
Ann. of Global Analysis and Geometry {\bf 28} (2005), 371--394.

\bibitem[20]{L} H.~B.~Lawson,
{\em Complete minimal surfaces in $S^3$},
Ann. Math. II Ser. {\bf92} (1970), 335--374.

\bibitem[21]{LPP} H.~L\"u, D.~Page and C.~N.~Pope,
{\em New inhomogeneous Einstein metrics on sphere bundles over
Einstein-K\"ahler manifolds}, (2004), 
hep-th/0403079.


\bibitem [22]{Mi} R.~Miyaoka, 
{\em The linear isotropy group of $G_2/SO(4)$, the Hopf fibering and
isoparametric hypersurfaces}, 
 Osaka J. of Math. {\bf 30} (1993), 179--202.

\bibitem [23]{Mu} M.~F.~M\"unzner, 
{\em Isoparametrische Hyperfl\"achen  in Sph\"aren I, II},
Math. Ann. {\bf 251}, (1980), 57--71, 
Math. Ann. {\bf 256} (1981), 215--232.

\bibitem [24]{OT} H.~Ozeki and M.~Takeuchi,
{\em On some types of isoparametric hypersurfaces in spheres, I,II},
T\^ohoku Math.J. {\bf 27} (1975), 515--559 and {\bf 28} (1976), 7--55.

\bibitem[25]{P}
B. Palmer,
{\em Hamiltonian minimality and Hamiltonian stability of Gauss maps},
Diff. Geom. and Appl. {\bf 7} (1997), 51--58.

\bibitem[26]{S} 
M.~B.~Stenzel,
{\em Ricci-flat metric on the complexification of compact rank one symmetric spaces},
Manuscripta Math. {\bf 80} (1993), 151--163.

\bibitem[27]{St} S.~Stolz,
{\em Multiplicities of Dupin hypersurfaces},
Invent. Math. {\bf138} (1999), 253--279.

\bibitem[28]{Th} G.~Thorbergsson,
{\em  A survey on isoparametric hypersurfaces and their generalizations}, 
Handbook of Differential Geometry Vol.1 
(2000), 963--995, Elsevier Science.

\bibitem[29]{TY} G.~Tian and S.~T.~Yau,
{\em Complete K\"ahler manifolds with zero Ricci curvature I,II}, 
J. A. M. S. {\bf 3} (1990), 579--609, and Invent. Math. {\bf 106}(1991), 27--60.
\end{thebibliography}
\end{document}